\documentclass[10pt,journal,twocolumn]{IEEEtran}
\IEEEoverridecommandlockouts
\usepackage{epsfig, cite,bm,algorithm,algorithmic,epstopdf,amsmath,subfigure,multirow,amssymb,amsfonts,amsthm,xcolor,caption,graphicx,textcomp,xcolor, subfigure,soul}
\usepackage[inline, shortlabels]{enumitem}
\setlist{itemjoin ={,\enspace},itemjoin* = { and\enspace}}
\def\QEDclosed{\mbox{\rule[0pt]{1.5ex}{1.5ex}}}
\usepackage[T1]{fontenc}
\usepackage{stfloats}
\allowdisplaybreaks[4]

\allowdisplaybreaks[4]
\usepackage{booktabs}
\soulregister{\cite}7
\soulregister{\ref}7

\begin{document}

\title{Flexible Rate-Splitting Multiple Access for Near-Field Integrated Sensing and Communications}

\author{Jiasi Zhou, Cong Zhou, Cheng Zeng, and Chintha Tellambura,~\IEEEmembership{Fellow,~IEEE}
\thanks{Jiasi Zhou is with the School of Medical Information and Engineering, Xuzhou Medical University, Xuzhou, 221004, China, (email: jiasi\_zhou@xzhmu.edu.cn). (\emph{Corresponding author: Jiasi Zhou}).}
\thanks{Cong Zhou is with the School of Electronic and Information Engineering, Harbin Institute of Technology, Harbin, 150001, China, (email: zhoucong@stu.hit.edu.cn).}
\thanks{Cheng Zeng is with the National Mobile Communications Research Laboratory, Southeast University, Nanjing, 210096, China, (email: czeng@seu.edu.cn).}
\thanks{ Chintha Tellambura is with the Department of Electrical and Computer Engineering, University of Alberta, Edmonton, AB, T6G 2R3, Canada (email: ct4@ualberta.ca).} 
\thanks{This work was supported by the national key research and development program of China (2020YFC2006600) and the Talented Scientific Research Foundation of Xuzhou Medical University (D2022027).}}
\maketitle
	
\begin{abstract} 
This letter presents a flexible rate-splitting multiple access (RSMA) framework for near-field (NF) integrated sensing and communications (ISAC). The spatial beams configured to meet the communication rate requirements of NF users are simultaneously leveraged to sense an additional NF target. A key innovation lies in its flexibility to select a subset of users for decoding the common stream, enhancing interference management and system performance. The system is designed by minimizing the Cram\'{e}r-Rao bound (CRB) for joint distance and angle estimation through optimized power allocation, common rate allocation, and user selection. This leads to a discrete, non-convex optimization problem. Remarkably, we demonstrate that the preconfigured beams are sufficient for target sensing, eliminating the need for additional probing signals. To solve the optimization problem, an iterative algorithm is proposed combining the quadratic transform and simulated annealing. Simulation results indicate that the proposed scheme significantly outperforms conventional RSMA and space division multiple access (SDMA), reducing distance and angle estimation errors by approximately 100\% and 20\%, respectively.
\end{abstract}


\section{Introduction} 
The integrated sensing and communications (ISAC) paradigm is poised to support many emerging applications, such as auto-driving and indoor positioning\cite{10135096}. It can enhance communication capacity and sensing resolution but will utilize extremely large-scale antenna arrays and operate at high frequencies \cite{10135096,10032057,huang2024space}. The Rayleigh distance (the boundary between the near and far field regions) increases as those trends occur, making near-field (NF) propagation dominant and waves becoming spherical. This offers an additional distance dimension, enabling simultaneous direction and distance estimation \cite{10579914}. In contrast, traditional far-field sensing primarily resolves angular information, while distance estimation typically requires significant bandwidth resources \cite{10579914}. Therefore, this NF capability opens up new ISAC possibilities. 

Recent NF communications advances have demonstrated that spatial beams preconfigured for legacy users can be repurposed to support additional users \cite{10315058,10129111}, significantly enhancing system throughput and connectivity. These works provide a new way of thinking about whether these preconfigured beams can be leveraged to sense additional targets, thereby enabling NF ISAC. However, spherical-wave propagation results in spotlight-like beam focusing, where beam energy is concentrated on a specific point \cite{10135096,10579914}. As a result, two key questions arise: (a) whether these preconfigured spotlight-like beams can be effectively used for target sensing and (b) whether additional probing signals are necessary. These critical questions form the basis of this work.

The coexistence of communication and sensing tasks inherently creates a propagation environment with interference signals, which must be effectively mitigated. Conventional NF and ISAC studies have primarily employed space-division multiple access (SDMA) \cite{10135096,10579914,10520715} or non-orthogonal multiple access (NOMA) \cite{10315058,10129111}, each addressing interference differently. SDMA treats it as additional noise but suffers from a performance plateau when interference levels become excessive. In contrast, NOMA decodes and removes strong interference signals, but its performance depends heavily on complex receiver designs and significant user channel differences. Both approaches lack flexibility, often resulting in suboptimal performance \cite{10486996}. Rate-splitting multiple access (RSMA) allows receivers to partially decode interference signals while tolerating residual interference \cite{10486996}. It thus provides greater versatility and robustness in managing interference. Current RSMA far-field ISAC studies \cite{10486996,10032141,10287099} require all users to decode a common stream, limiting performance gains as the weakest user's channel constrains the common rate. This creates the need for a more advanced RSMA framework.

To our knowledge, even traditional RSMA-enabled NF ISAC (\emph{i.e.}, without selecting rate-splitting users) remains largely unexplored. This paper addresses this gap by proposing a flexible RSMA scheme for NF ISAC, with a careful selection of rate-splitting users. The primary objective is to explore how preconfigured beams can be utilized to sense an additional NF target. It is first proven that no extra probing signal is necessary for target sensing. Subsequently, an iterative optimization algorithm is developed, based on the quadratic transform and simulated annealing, to minimize the Cram\'{e}r-Rao bound (CRB)  estimation. This algorithm jointly optimizes power allocation, common rate allocation, and user selection. Simulation results indicate that our proposed scheme offers significant performance gains over several baselines.
\vspace{-0.5cm}
\section{System model and problem formulation} \label{Section II}
\begin{figure}[tbp]
\centering
\includegraphics[scale=0.7]{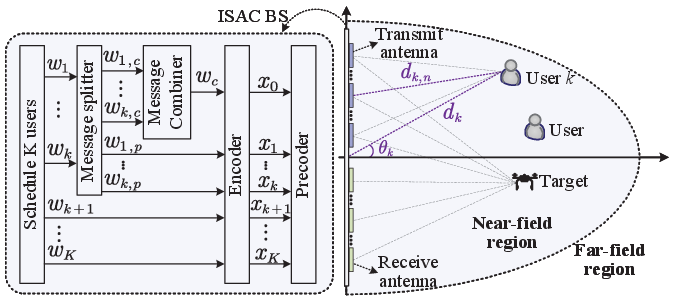}
\caption{Flexible RSMA-enabled NF ISAC.}
\vspace{-0.4cm}
\label{fig:system}
\end{figure}
Fig.~\ref{fig:system} considers a flexible RSMA-enabled NF ISAC network, which comprises a dual-functional base station (BS), $K>1$ single-antenna NF users, and one sensing target. The BS utilizes a uniform linear array (ULA) consisting of $N_t$-transmit and $N_r$-receive antennas with an antenna spacing of $d$. Therefore, the antenna apertures of transmit and receive ULA are respectively $D_t=(N_t-1)d$ and $D_r=(N_r-1)d$, resulting in the Rayleigh distance $Z_i=\frac{2D^2_i}{\lambda}$ for $\forall i\in\{t,r\}$, where $\lambda$ is the signal wavelength.
\vspace{-0.5cm}
\subsection{NF communication and sensing channel models}
The coordinate of the $n$-th transmit antenna is defined as $\mathbf{s}_n=\left(0, nd\right)$, where $n\in\mathcal{N}_t=\{1,\dots,N_t\}$. 
Let $d_k$ and $\theta_k$ denote the distance and angle of user $k$, so its coordinate is $\mathbf{d}_k=\left(d_k\cos\theta_k,d_k\sin\theta_k\right)$. The distance between the $n$-th transmit antenna and the $k$-th user is given by 
\begin{equation}
d_{k,n}=||\mathbf{d}_k-\mathbf{s}_n|| = \sqrt{d^2_k+( nd)^2-2ndd_k\sin\theta_k}.
\end{equation}
The second-order Taylor expansion yields  $d_{k,n}\approx d_k-\delta_{k,n}$, where $\delta_{k,n}=nd\sin\theta_k-(nd)^2\cos^2\theta_k/2d_k$. Consequently, the corresponding NF channel is $h_{k,n}= \tilde\beta_k e^{-j\frac{2\pi}{\lambda}(d_k-\delta_{k,n})}$, where $\tilde\beta_k$ is the free space path loss. Specifically, $\tilde \beta_k=\frac{c}{4\pi fd_k}$, where $f$ and $c$ are the carrier frequency and speed of light, respectively. Therefore, the NF channel vector $\mathbf{h}_k\in\mathbb{C}^{N_t\times 1}$ between the BS and the $k$-user can be modeled as
\begin{equation}
\mathbf{h}_k= \beta_k\big[e^{j\frac{2\pi}{\lambda}\delta_{k,1}},\dots,e^{j\frac{2\pi}{\lambda}\delta_{k,N_t}}\big]^T=\beta_k\mathbf{a}\left(d_k,\theta_k\right),
\end{equation}
where $\beta_k=\tilde\beta_ke^{-j\frac{2\pi}{\lambda}d_{k}}$ and $\mathbf{a}\left(d_k,\theta_k\right)$ is the NF array response vector. 

For sensing the target, the BS sends probing signals to it and then gathers the echo signals. Thus,   the round-trip channel (i.e., sensing channel) must be considered \cite{10135096,10579914}. To describe it, let  $d_s$ and $\theta_s$ denote the distance and angle of the sensing target. Similar to NF channel modeling, the sensing channel matrix $\mathbf{G}\in\mathbb{C}^{N_r\times N_t}$ can be modeled as 
\begin{equation}
\mathbf{G}=\beta_s\mathbf{a}_r\left(d_s,\theta_s\right)\mathbf{a}^T_t\left(d_s,\theta_s\right),
\end{equation}
where $\beta_s$ captures the round-trip path loss. $\mathbf{a}_r\left(d_s,\theta_s\right)\in\mathbb{C}^{N_r\times 1}$ and $\mathbf{a}_t\left(d_s,\theta_s\right)\in\mathbb{C}^{N_t\times 1}$ denote the receive and transmit NF array response vector, respectively.
\vspace{-0.5cm}
\subsection{Signal model and problem formulation}
Unlike traditional RSMA, our scheme selects a subset of users to decode the common stream. To be specific, $K$ users are divided into two separate groups, \emph{i.e.}, group 1 and group 2 are respectively denoted by $\mathcal{K}_1=\{k|s_k=1\}$ and $\mathcal{K}_2=\{k|s_k=0\}$ for $\forall k\in\mathcal{K} =\{1,\dots,K\}$, where $s_k\in\{0,1\}$ is the user selection indicator. Message $s_k$ for user $k$ in $\mathcal{K}_1$ is split into common and private parts. All common parts in $\mathcal{K}_1$ are encoded into a common stream $x_0$, while each private message is encoded into a private stream $x_k$. The message for user $\tilde k$ in $\mathcal{K}_2$ is directly encoded into a stream $x_{\tilde k}$. These streams are mutually independent and linearly precoded by preconfigured $\mathbf{p}_0\in\mathbb C^{N_t\times 1}$ and $\mathbf{p}_k\in\mathbb C^{N_t\times 1}$. Herein, the zero-forcing principle is employed for private streams, \emph{i.e.}, $\mathbf{P}_p=[\mathbf{p}_1,\dots,\mathbf{p}_K] = \mathbf{H}\left(\mathbf{H}^H\mathbf{H}\right)^{-1}\mathbf{Q}$, where $\mathbf{H}=[\mathbf{h}_1,\dots,\mathbf{h}_K]$ and $\mathbf{Q}\in\mathbb{C}^{N_t\times N_t}$ is a diagonal matrix to ensure power normalization; consequently,  $[\mathbf{Q}]_{n,n} =\big[\left(\mathbf{H}^H\mathbf{H}\right)^{-1}\big]^{-\frac{1}{2}}_{n,n}$. Moreover, since the common rate depends on the user with the worst channel quality,  set $\mathbf{p}_0=\mathbf{h}_{k'}$, where $k'=\arg\min_k\left\{||\mathbf{h}_k||\right\}$ for $\forall k\in\mathcal{K}_1$. 
Precoder optimization lies beyond this letter's scope, primarily leveraging preconfigured beams to accommodate an additional sensing target. Nonetheless, this remains a valuable topic for future research. 

The BS sends communication and sensing signals over a coherent time block of length $T$, during which the channel and target parameters remain roughly constant. Therefore, the transmitted signal at time slot $t$ can be written as 
$\mathbf{x}(t)=\sum_{k=0}^{K}\mathbf{p}_{k}\left(\sqrt{P_k}x_k(t) +\sqrt{\tilde P_{k}}\tilde x(t) \right)$\footnote{This paper focuses on how to repurpose preconfigured beamformers, so the probing signal is not equipped with a dedicated beamformer.},
where $\tilde x(t)$ is the probing signal. $P_k$ and $\tilde{P}_k$ are the transmit power allocated to the user and probing signal on the $k$-th beam. The received signal at user $k$ is $
y_k(t)=\mathbf{h}^H_k\mathbf{x}(t) + n_k(t)$,
where $n_k(t)\sim \mathcal{CN}\left(0,\sigma^2_k\right)$ is the additive white Gaussian noise (AWGN) term. 

To detect the desired message, user $k$ in $\mathcal{K}_1$ first decodes the common stream. Based on the known precoders,  the signal-to-interference-plus-noise ratio (SINR) can be written as
\begin{equation}
\gamma_{k,c}=\frac{h_{k,0}P_0}{h_{k,0}\tilde P_0+h_{k,k}\left(P_k+\tilde P_k\right) + \sigma^2_k}.\label{common_rate}
\end{equation}
where $h_{k,i}=\left|\mathbf{h}^H_{k}\mathbf{p}_i\right|^2$ for $i\in\{0,k\}$. To ensure that $x_0$ can be detected by all users in $\mathcal{K}_1$, the common rate should not exceed $R_c= \min_{\forall k\in\mathcal{K}_1}\log \left(1+\gamma_{k,c}\right)$. Moreover, since $R_c$ is shared by all users in  $\mathcal{K}_1$,  one has  $\sum_{k=1}^{K}s_kR_{k,c} \leq R_c$, where $R_{k,c}$ is the $k$-th user's common rate. After removing the common stream, users in $\mathcal{K}_1$ decode their desired private stream. In contrast, users in $\mathcal{K}_2$ directly decode their private stream. To save pages, this paper merges the SINRs of all users decoding private streams into one expression, which is as follows,
\begin{equation}
\gamma_{k,p}=\frac{h_{k,k}P_k}{h_{k,0}\left((1-s_{k})P_0+\tilde P_0\right)+h_{k,k}\tilde P_k + \sigma^2_k},\label{private_common}
\end{equation}
where $k\in\mathcal{K}$. As a result, the transmit rate of the $k$-th user is $R_k=s_k R_{k,c}+\log\left(1+\gamma_{k,p}\right)$.

The received echo signal at the BS can be written as 
\begin{equation}
\mathbf{y}_s\left(t\right) = \mathbf{G}\mathbf{x}\left(t\right) + \mathbf{H}_{SI}\mathbf{x}\left(t\right) + \mathbf{n}\left(t\right),
\end{equation}
where $\mathbf{H}_{SI}\in\mathbb{C}^{N_r\times N_t}$ is self-interference channel from transmit to receive ULA and $\mathbf{n}\left(t\right)\sim \mathcal{CN}\left(0,\sigma^2\mathbf{I}_{N_r}\right)$ is the AWGN. Similar to \cite{10579914}, assuming perfect self-interference cancellation, the received echo signal over $T$ coherent time slot is $\mathbf{Y}_s=\mathbf{G}\mathbf{X}+\mathbf{N}$, where $\mathbf{Y}_s=\big[\mathbf{y}_s\left(1\right),\dots,\mathbf{y}_s\left(T\right)\big]$ and $\mathbf{N}=\big[\mathbf{n}\left(1\right),\dots,\mathbf{n}\left(T\right)\big]$. This paper utilizes CRB as the sensing performance metric. The CRB matrix is given by 
\begin{equation}
\mathrm{CRB}\left(\mathbf{R},\mathbf{G},\sigma^2\right) = \left(\mathbf{F}_{11}-\mathbf{F}_{12}\mathbf{F}^{-1}_{22}\mathbf{F}^T_{12}\right)^{-1},\label{CRB}
\end{equation}
where 
\begin{subequations}\label{FIM}
	\begin{align}
 &\mathbf{F}_{11}= \frac{2|\beta_s|^2T}{\sigma^2}\mathrm{Re}\left(\begin{bmatrix} F_{\theta_s\theta_s} & F_{\theta_s d_s} \\ F_{\theta_s d_s} & F_{d_sd_s} \end{bmatrix}\right),\\
&\mathbf{F}_{1,2}=\frac{2T}{\sigma^2}\mathrm{Re}\left(\begin{bmatrix} \beta^*_s\mathrm{Tr}(\mathbf{\tilde G}\mathbf{R}\mathbf{G}^H_{\theta_s}) \\ \beta^*_s\mathrm{Tr}(\mathbf{\tilde G}\mathbf{R}\mathbf{G}^H_{x}) \end{bmatrix}\begin{bmatrix} 1,j \end{bmatrix}\right),\\
 &\mathbf{F}_{2,2}=\frac{2T}{\sigma^2}\mathbf{I}_2\mathrm{Tr}\left(\mathbf{\tilde G}\mathbf{R}\mathbf{\tilde G}^H\right).
	\end{align}
\end{subequations}
The derivation of the CRB matrix is similar to Appendix B in  \cite{10579914}. We omit the detailed derivation due to the limited pages. In equation (\ref{FIM}), $F_{xy}=\mathrm{Tr}\left(\mathbf{G}_{y}\mathbf{R}\mathbf{G}^H_{x}\right)$, $ \mathbf{ \tilde G}= \mathbf{a}_r\left(d_s,\theta_s\right)\mathbf{a}^T_t\left(d_s,\theta_s\right)$, $\mathbf{G}_{d_s}=\frac{\partial{\mathbf{\tilde G}}}{\partial d_s}$, $\mathbf{G}_{\theta_s}=\frac{\partial{\mathbf{\tilde G}}}{\partial \theta_s}$ and $\mathbf{R}=\mathbb{E}\left[\mathbf{x}(t)\mathbf{x}^H(t)\right] =\sum_{k=0}^{K}\mathbf{p}_{k}\mathbf{p}^H_{k}\left(P_k +\tilde P_{k} \right)$. Consequently, we have $\epsilon^2_{\theta_s}\geq [\mathrm{CRB}]_{1,1}$ and $\epsilon^2_{d_s}\geq [\mathrm{CRB}]_{2,2}$.

This paper aims to minimize the trace of the CRB matrix by jointly optimizing power allocation, common rate allocation, and user selection. This problem is formulated as
\begin{subequations}\label{linear_p}
	\begin{align}
&\min_{P_{k},\tilde P_{k},R_{k,c},s_k} \mathrm{Tr}\left(\mathrm{CRB}\left(\mathbf{R},\mathbf{G},\sigma^2\right)\right),\label{ob_a}\\
	\text{s.t.}\quad
	&\sum_{k=0}^{K} \left(P_k +\tilde P_{k} \right)\leq P_{max},\label{ob_b}\\&
 R_k\geq R_{th}, \quad \forall k,\label{ob_c}\\
 &\sum_{k=1}^{K}s_kR_{k,c} \leq R_c,\label{ob_d}\\
 &R_{k,c} \geq 0,\label{ob_e}
	\end{align}
\end{subequations}
where $P_{max}$ and $R_{th}$ are maximum transmit power and quality of services (QoS) thresholds, respectively. 

Problem (\ref{linear_p}) is a discrete, non-convex optimization problem, posing two significant challenges. First, determining the optimal user selection demands an exhaustive search, which is computationally prohibitive and challenging to implement in practice. Second,  the objective function and decoding rate exhibit non-convexity and non-smoothness, complicating solution approaches in primal and dual domains due to the unknown duality gap. Consequently, finding a globally optimal solution is mathematically intractable.  

\vspace{-0.5cm}
\section{Algorithm design and properties analysis}\label{Section III}
This section rigorously demonstrates that no additional probing signal is required for target sensing. The problem (\ref{linear_p}) is then divided into two sub-problems: 1) power and common rate allocation, and 2) user selection. An iterative optimization algorithm based on the quadratic transform and simulated annealing is developed to address these sub-problems. Additionally, the optimality, convergence, and complexity of the proposed algorithms are discussed.
\vspace{-0.5cm}
\subsection{Is an extra probing signal required for target sensing?} 
To make the problem (\ref{linear_p}) more tractable, we first determine whether the extra probing signal is needed or not. 

{\bf\emph{Proposition 1}}: Under the optimal resource allocation and user selection, the extra probing signal is not required for target sensing,  \emph{i.e.}, $\tilde P^*_k=0$ for $\forall k\in\tilde{\mathcal{K}}=\{0,1,\dots, K\}$.

{\emph{Proof}}: Proposition 1 is proved via contradiction. Assuming that there is a $k\in\tilde{\mathcal{K}}$, the optimal ${\tilde P}^*_k \neq 0$. Let $P^*_k$ denote the corresponding transmit power for the $k$-th communication user. Then, we update $P_k = P^*_k + {\tilde P}^*_k$ and ${\tilde P}_k = 0$ but keep other variables unchanged. Plugging the latest $P_k$ and ${\tilde P}_k$ into (\ref{common_rate}), (\ref{private_common}), and (\ref{CRB}), one can derive that $\gamma_{k,c}$ and $\gamma_{k,p}$ are increasing while the SINR of other communication users and the CRB of sensing target remain static. This indicates that network performance can be enhanced after updating the resource allocation strategy, which contradicts the primary assumption.  Proposition 1 is thus proved.\hfill \QEDclosed

Proposition 1 shows that preconfigured NF beams are sufficient for target sensing. Furthermore, with Proposition 1, we recast (\ref{common_rate}), (\ref{private_common}), (\ref{ob_a}) and (\ref{ob_b}) as follows,
\begin{subequations}\label{new-sinr}
	\begin{align}
&\gamma_{k,c}=\frac{h_{k,0}P_0}{h_{k,k}P_k + \sigma^2_k},\quad k\in\mathcal{K}_1,\\
&\gamma_{k,p}=\frac{h_{k,k}P_k}{(1-s_{k})h_{k,0} + \sigma^2_k},\quad k\in\mathcal{K},\\
&\mathrm{Tr}\left(\mathrm{CRB}\left(\mathbf{R},\mathbf{G},\sigma^2\right)\right)=\mathrm{Tr}\left(\mathrm{CRB}\left(\tilde{\mathbf{R}},\mathbf{ G},\sigma^2\right)\right),\\
&\sum_{k=0}^{K} P_k \leq P_{max},\label{power_constraint}
	\end{align}
\end{subequations}
where $\tilde{\mathbf{R}} = \sum_{k=0}^{K}\mathbf{p}_{k}\mathbf{p}^H_{k}P_k$.
Although $\tilde P^*_k=0$ is known, a direct solution for problem (\ref{linear_p}) is still elusive. To attack this challenge, it is divided into two sub-problems. These two sub-problems and their solution are as follows. 
\vspace{-0.5cm}
\subsection{Power and common rate allocation sub-problem}
With the known $s_k$ and optimal $\tilde P^*_k$, problem (\ref{linear_p}) still present coupled power allocation and fractional SINR. Surrogate optimization is utilized to address this issue.  This method requires constructing accurate and easily optimized surrogates for the objective and constraints. Herein,  the quadratic transform approach is utilized to construct surrogates, which decouples the fractional SINR into a difference of two terms. Theorem 2 in \cite{shen2018fractional} motivates our Claim 1.

{\bf\emph{Claim 1}}: For function $f\left(y,p\right) = 2y\sqrt{s(p)}-y^2 I(p)$ for any $s(p)\geq 0$ and $I(p)> 0$, we have
\begin{equation}
s(p)I^{-1}(p) =\max_{y>0}f\left(y,p\right)
\label{Surrogate}.
\end{equation} 
The optimal solution to  $\max_{y>0}f\left(y\right)$ is $y^*=\sqrt{s(p)}I^{-1}(p)$.

\emph{Proof}: Please see \cite{shen2018fractional} for the detailed proof.\hfill \QEDclosed

Based on Claim 1, the surrogate function for $\gamma_{k,c}$ and $\gamma_{k,p}$ can be respectively written as
\begin{subequations}\label{surrogate}
	\begin{align}
&f\left(P,y_{k,c}\right)=2y_{k,c}\sqrt{h_{k,0}P_0}-y^2_{k,c}\left(h_{k,k}P_k + \sigma^2_k\right),\\
 &f\left(P,y_{k,p}\right)=2y_{k,p}\sqrt{h_{k,k}P_k}-y^2_{k,p}\left((1-s_k)h_{k,0}P_0 + \sigma^2_k\right).
	\end{align}
\end{subequations}

The non-convex CRB matrix is another obstacle in solving the problem (\ref{linear_p}). To attack this challenge, we introduce an auxiliary matrix and recast the objective function into an equivalent but more tractable convex form. After removing the minimum operator in decoding the common stream, problem (\ref{linear_p}) can be recast as
\begin{subequations}\label{linear_p3}
	\begin{align}
&\min_{\mathcal{Q}_1,\mathcal{Q}_2} \mathrm{Tr}\left(\mathbf{U}^{-1}\right),\label{ob_a3}\\
	\text{s.t.}~
 &\begin{bmatrix} \mathbf{F}_{11} -\mathbf{U}& \mathbf{F}_{12} \\ \mathbf{F}^T_{12} & \mathbf{F}_{22} \end{bmatrix}\succeq\mathbf{0},\label{ob_b3}\\
 &\sum_{k=1}^{K}s_kR_{k,c} \leq \log\left(1+f\left(P,y_{k,c}\right)\right),\forall k\in\mathcal{K}_1,\label{ob_c3}\\
  & s_kR_{k,c}+\log\left(1+f\left(P,y_{k,p}\right)\right)\geq R_{th},\forall k\in\mathcal{K},\label{ob_d3}\\
 &\mbox{(\ref{power_constraint}), (\ref{ob_e})},
	\end{align}
\end{subequations}
where $\mathcal {Q}_1=\{P_{k},R_{k,c},\mathbf{U}\succeq\mathbf{0}\}$ and $\mathcal {Q}_2=\{y_{k,c},y_{k,p}\}$. Observe that all constraints remain convex sets when $\mathcal{Q}_2$ is fixed, which can be solved by CVX. Besides, when $\mathcal{Q}_1$ is specified, the closed-form solution of the optimal $\mathcal{Q}_2$ can be derived via Claim 1. Hence, we solve the problem (\ref{linear_p3}) alternately over $\mathcal{Q}_i$ while keeping $\mathcal{Q}_j$ at its previous value, where $i,j\in\{1,2\}$ and $i\neq j$. Alg.~\ref{Alg.1} presents the outline of our proposed iterative algorithm. Here are its crucial properties.
\begin{itemize}
\item  \emph{Convergence and Optimality}:
Given an arbitrary feasible $\mathcal{Q}^{(1)}_1$, Alg.~\ref{Alg.1} always outputs the optimal solution in lines 3 and 4. It advances toward the most recent feasible point after each iteration, so the objective value remains either stable or decreases. Meanwhile. since the sensing performance is lower bounded by a finite value, we thus deduce that Alg.~\ref{Alg.1} converges within several iterations to a stationary point at least.

\item \emph{Complexity}: The main load comes from solving $\mathcal{Q}_1$ via CVX.  Using the conventional interior method, its computational complexity is $\mathcal O(V^{3.5})$, where $V$ is the number of variables. Thus, the complexity is  $\mathcal O\left(\epsilon_1\left(2K+4\right)^{3.5}\right)$, where $\epsilon_1$ is the iteration numbers until Alg.~\ref{Alg.1} ends. 
\end{itemize}
\begin{algorithm}[t]
	\caption{Quadratic transform-based iteration algorithm}
	\begin{algorithmic}[1]\label{Alg.1}
        \STATE Initialize $\mathcal{Q}_1^{(1)}$ and iteration index $i=1$.
        \REPEAT
		\STATE  Calculate $\mathcal{Q}_2^{(i)}$ under fixed $\mathcal{Q}_1^{(i)}$ based on Claim 1.
		\STATE  Solve problem (\ref{linear_p3}) under fixed $\mathcal{Q}_2^{(i)}$  and  output $\mathcal{Q}^{(i+1)}_1$.
		\STATE Update $i=i+1$.
		\UNTIL{The decrease of $\mathrm{Tr}\left(\mathbf{U}^{-1}\right)$ is less than $\delta_1=10^{-3}$.}	
		\STATE Obtain the solution $\left\{P_k,R_{k,c},\mathbf{U}\right\}$ with given $\mathbf{s}$.
	\end{algorithmic}
\end{algorithm}
\vspace{-0.5cm}
\subsection{User selection sub-problem}
Given the user selection $\mathbf{s}=\{s_1,\dots,s_k\}$, one can get the corresponding optimal CRB matrix via Alg.~\ref{Alg.1}. Naturally, we can find the globally optimal solution by exhaustively searching all possible combinations with Alg.~\ref{Alg.1}. Nevertheless, the complexity of exhaustive searching $\mathcal{O}\left(2^K\right)$, which is prohibitive. This limitation calls for a low-complexity user selection approach. For this, simulated annealing is deployed to search for a good user selection solution. This scheme has a probabilistic nature, which allows it to escape the local optimum via exploration. Alg.~\ref{Alg.2} presents the overall process for solving problem (\ref{linear_p}). It initializes with a random user selection strategy $\mathbf{s}$. In each step, it is tested whether changing $s_k$ to $1-s_k$ can reduce estimation error. If it holds,  user selection is updated to  $\tilde{\mathbf{s}}=\{s_1,\dots,1-s_k,\dots,s_K\}$. Otherwise, it is  still changed  with a probability $
\mathbb{P}=\exp\left(\frac{V\left({\mathbf{s}}\right)-V\left(\tilde{\mathbf{s}}\right)}{\delta}\right)\in(0,1]$,
where $V\left(\tilde{\mathbf{s}}\right)$ and $V\left({\mathbf{s}}\right)$ denote the objective value when user selection strategy is $\tilde{\mathbf{s}}$ and $\mathbf{s}$. $\delta$ is a large positive "\emph{temperature}" parameter. In general, $\delta=\rho\delta$ aims to accelerate the convergence rate, where $0<\rho<1$ is the decaying rate. Several crucial properties are discussed as follows.

\begin{itemize}
\item\emph{Convergence and optimality}: 
The overall algorithm integrates the simulated annealing approach with Alg.\ref{Alg.1}. As the parameter $\delta$ decreases during the simulated annealing process, the probability $\mathbb{P}$ also decreases, aiding in the convergence to the optimal solution \cite{10414053}. Additionally, since Alg.\ref{Alg.1} has been proven to converge in Section \ref{Section III}-B, the overall algorithm is guaranteed to converge after a finite number of iterations.

\item\emph{Complexity}: Most of  complexity arises  from Alg.~\ref{Alg.1}. Assuming Alg.~\ref{Alg.2} converges after $\epsilon_2$ iterations, its complexity is $\mathcal O\left(\epsilon_1\epsilon_2\left(2K+4\right)^{3.5}\right)$.
\end{itemize}

\begin{algorithm}[t]
	\caption{Overall algorithm for solving problem (\ref{linear_p})}
	\begin{algorithmic}[1]\label{Alg.2}
        \STATE Initialize user selection strategy $\mathbf{s}^{(0)}$, optimal sensing performance $V^*=0$ and iteration index $t=0$.
        \STATE Initialize sensing performance $\tilde V=V\left({\mathbf{s}}^{(0)}\right)$ via Alg.~\ref{Alg.1}.
        \REPEAT
        \STATE Set $\tilde {\mathbf{s}}=\mathbf{s}^{(t)}$ and update $\tilde {s}_k = 1 - \tilde s_k$ and $\mathbf{p}_0$, where $k={\rm{mod}}\left(t,K\right)+1$. Then, obtain  $V\left(\tilde{\mathbf{s}}\right)$ via Alg.~\ref{Alg.1}.
		\IF{ $V\left(\tilde{\mathbf{s}}\right)<\tilde V$ }
		\STATE Update ${\mathbf{s}}^{(t+1)}=\tilde{\mathbf{s}}$ and $\tilde V = V\left(\tilde{\mathbf{s}}\right)$.
        \IF{$V\left(\tilde{\mathbf{s}}\right)<V^*$}
        \STATE Update $V^* = V\left(\tilde{\mathbf{s}}\right)$
        \ENDIF
        \ELSE
        \STATE Update ${\mathbf{s}}^{(t+1)}=\tilde{\mathbf{s}}$ and $\tilde V = V\left(\tilde{\mathbf{s}}\right)$ with  probability $\mathbb{P}$. Keep ${\mathbf{s}}^{(t+1)} ={\mathbf{s}}^{(t)}$ with probability $1-\mathbb{P}$.
		\ENDIF
		\STATE Update $t=t+1$ and $\delta = \rho \delta$.
		\UNTIL{The change of $V^*\leq \delta_2=10^{-3}$ for the last $K$ iterations.}
		\STATE Output the optimal trace of the CRB matrix. 
	\end{algorithmic}
\end{algorithm}

\vspace{-0.5cm}
\section {Simulation results}
Next, we provide numerical results to evaluate the proposed transmit scheme and algorithm. Unless stated otherwise, the simulation parameters are as follows: The BS is equipped with $N_t=128$ and $N_r=64$ antennas operating at a frequency of $f_c=30$~GHz. The antenna space is $d=0.5\lambda$, so $Z_t=80.6$~m and $Z_i=20.5$~m. The polar coordinates of the sensing target are $(15~\mathrm{m},45^{\circ})$. $K=20$ users are randomly generated within the distance from $15$~m to $25$~m.  The maximum transmit power at the BS and background noise power are $P_{max}=30$~dBm and $\sigma^2=-80$~dBm, respectively. The QoS of NF users is $R_{th}=3$~bps/Hz. The Alg.~\ref{Alg.2} parameters are set as $\delta=20$ and $\rho=0.9$. These parameters are primarily taken from \cite{10135096,10414053}. In the simulation figures, "RCRB" denotes the root of CRB.

Over 200 channel realizations,  our scheme and algorithm (labeled as {\bf{FRS}}) are compared against the two benchmarks:
\begin{itemize}
\item {\bf{RS}}: The common stream is decoded by all users, \emph{i.e.}, $\mathcal{K}_1= \mathcal{K}$ and $\mathcal{K}_2= \emptyset$. 
\item {\bf{SDMA}}: Each user's message is encoded into a private stream, \emph{i.e.}, $\mathcal{K}_1= \emptyset$ and $\mathcal{K}_2= \mathcal{K}$. NOMA reduces to SDMA when the zero-forcing technique is adopted.
\end{itemize}

\begin{figure}[tbp]
	\centering
	\subfigure[RCRB of distance versus users]{
		\begin{minipage}[t]{0.46\linewidth}
			\centering\includegraphics[width = 1.63in]{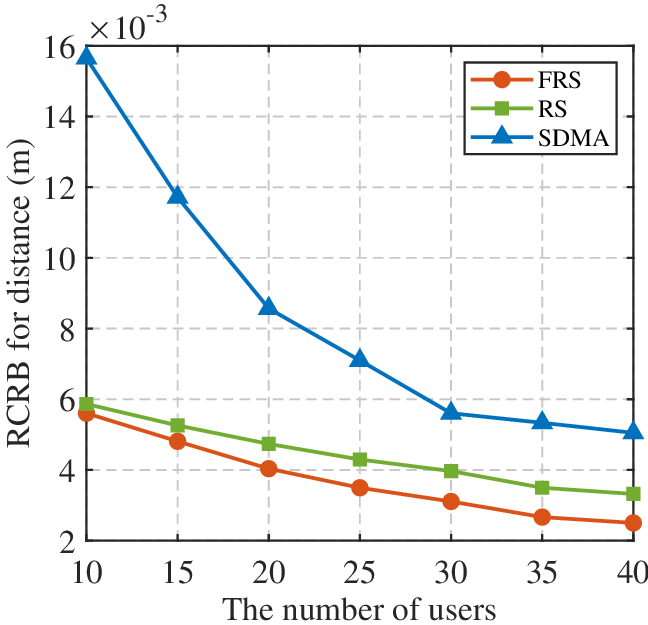}
		\end{minipage}
	}
	\subfigure[RCRB of angle versus users]{
		\begin{minipage}[t]{0.46\linewidth}
			\centering\includegraphics[width = 1.58in]{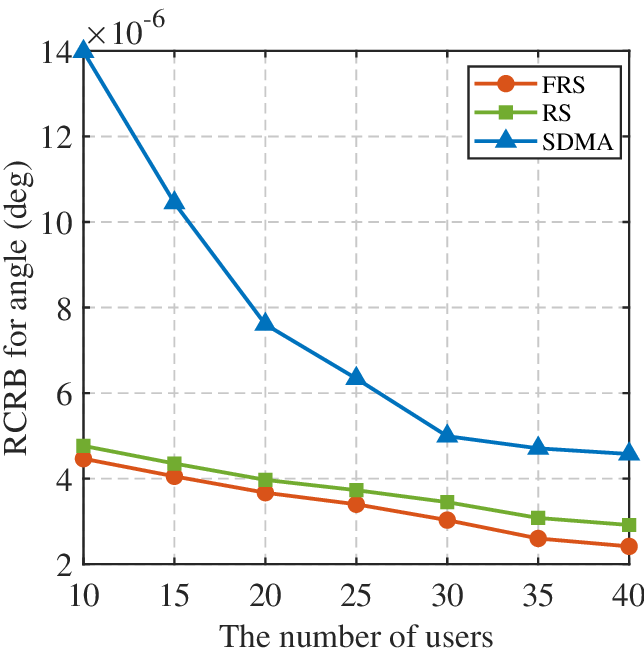}
		\end{minipage}
	}
	\centering
	\caption{RCRB versus the number of users}
	\label{Fig_user}
    \vspace{-0.3cm}
\end{figure}

\begin{figure}[tbp]
	\centering
	\subfigure[RCRB of distance versus QoS]{
		\begin{minipage}[t]{0.46\linewidth}
			\centering\includegraphics[width = 1.58in]{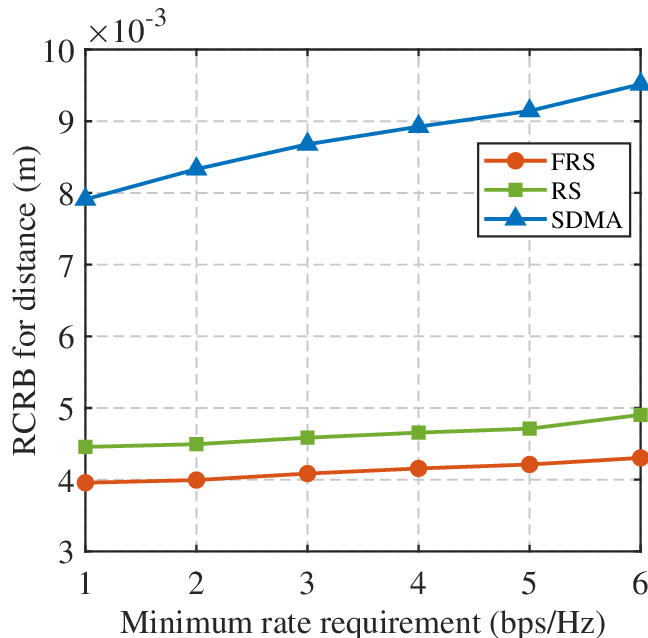}
		\end{minipage}
	}
	\subfigure[RCRB of angle versus QoS]{
		\begin{minipage}[t]{0.46\linewidth}
			\centering\includegraphics[width = 1.58in]{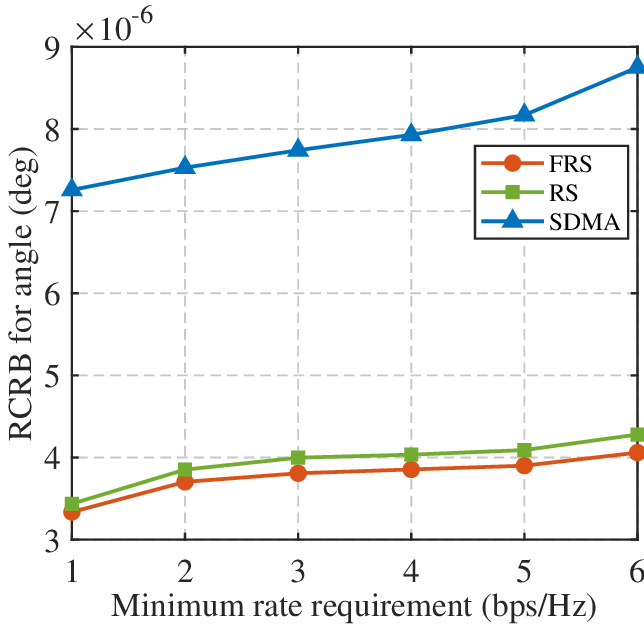}
		\end{minipage}
	}
	\centering
	\caption{RCRB versus the QoS of users}
	\label{Fig_QoS}
    \vspace{-0.3cm}
\end{figure}

Fig.~\ref{Fig_user} illustrates the RCRB for distance and angle as a function of the number of users, revealing three key insights. First, the RCRB decreases as the number of users increases, which is counterintuitive. However, this occurs because the number of non-parallel precoder vectors grows linearly with the number of users, broadening the vector space they can represent and enhancing the potential for target sensing. Second, both conventional RSMA and our proposed flexible RSMA (i.e., FRS) outperform SDMA, particularly when scheduling a small number of users, underscoring its effectiveness in managing interference. Third, our proposed solution, FRS,  surpasses traditional RSMA, with the RCRB gap increasing as the number of users increases. This improvement stems from the ability of FRS to selectively choose a subset of users to decode the common stream. As the number of users grows, the solution space for user selection expands, and the simulated annealing approach effectively identifies the optimal rate-splitting users.

Fig.~\ref{Fig_QoS} presents the RCRB for distance and angle versus QoS requirements. As expected, the RCRB of each scheme decreases as the minimum rate increases. This occurs because some precoder vectors, poorly aligned for sensing, require more power to meet the minimum rate, thus adversely affecting the RCRB. Furthermore, compared to SDMA and traditional RSMA, FRS  reduces the estimation error for distance and angle by approximately 100\% and 20\%, respectively, demonstrating that user selection enhances target sensing capabilities. Additionally, the estimation error of RSMA increases slower than with SDMA, further highlighting RSMA's effectiveness in managing interference.

\section{Conclusion}
This paper presents a flexible RSMA scheme for NF ISAC that leverages preconfigured beams to support additional target sensing. The problem of CRB minimization, being discrete and non-convex, is addressed by proving that no additional probing signal is required to enhance sensing performance. An iterative optimization algorithm is then developed to jointly optimize power allocation, common rate allocation, and user selection. Simulations demonstrate that our scheme and algorithm achieve significant gains over several baselines. Future research directions include exploring hybrid precoding designs for NF ISAC and expanding our algorithm to accommodate imperfect CSI and multi-target sensing.

	\ifCLASSOPTIONcaptionsoff
	\newpage
	\fi
	
	\bibliographystyle{IEEEtran}
	\bibliography{references}
	
\end{document}